\documentstyle[aps,preprint,tighten]{revtex}

\begin{document}

\draft


\title{Electric quadrupole moments of the decuplet
       baryons in the Skyrme model}

\author{Yongseok Oh}

\address{Department of Physics, National Taiwan University,
         Taipei, Taiwan 10764, Republic of China}


\maketitle

\widetext

\begin{abstract}
The electric quadrupole moments of the decuplet baryons are calculated
in the bound state approach to the SU(3) Skyrme model. In this approach, all
the quadrupole moments of the decuplets are found to be proportional
to the third component of the baryon isospin. Contrary to the SU(3)
collective model, the kaonic contribution is as important as the pionic
one. The transitional quadrupole moments of hyperons are also
predicted.
\end{abstract}


\narrowtext

In the quark model, the tensor force in the inter-quark hyperfine
interaction \cite{DGG} leads to the $D$-wave admixture in the
baryon wave functions, so that the ground state baryons are slightly
deformed. As a result, the tensor force induces a small violation of
Becchi-Morpurgo selection rule \cite{BM} that the decay $\Delta \to
\mbox{N} \gamma$ is pure $M1$ transition, and also leads to
non-vanishing electric quadrupole moments of the decuplet baryons.
Following the simple calculation in the oscillator model which
predicts measurable $\Omega^-$ quadrupole moment \cite{GZ}, there are
a number of publications concerning the quadrupole moments of the
non-strange and strange decuplet baryons in various phenomenological
models \cite{GD,IKK,Ric,DG,Kri,LG,KG,LDW,BSS,KS} and the results are
highly model-dependent.

Recently, Leinweber {\it et al.} \cite{LDW} studied the quadrupole
moments in the lattice simulation of QCD and Butler {\it et al.}
\cite{BSS} used heavy baryon chiral perturbation theory \cite{JM} for
this study. Based on the ``slow rotator'' approach to the SU(3)
Skyrme model \cite{SW}, Kroll and Schwesinger \cite{KS} investigated
this problem in connection with the strangeness content of the proton.
In the flavor SU(3) limit, the Skyrme model predicts maximum value of
the strangeness content of the proton, $\langle s \bar s \rangle_p =
\frac{7}{30}$, whereas $\langle s \bar s \rangle_p \to 0$ in
the strong symmetry breaking limit. In Ref. \cite{KS}, the authors
claimed that the quadrupole moments are proportional to the baryon
charge in the SU(3) limit and to the isospin of the baryon in the
symmetry breaking limit. Their numerical results obtained with the
physical kaon mass are somewhat different from those of Refs.
\cite{LDW,BSS}.

In this paper, we study the electric quadrupole moments of the decuplet
baryons in the bound state approach to the SU(3) Skyrme model. This
approach suggested by Callan and Klebanov \cite{CK} is based on the
observation that the $s$ quark is much heavier than the light $(u,d)$ quarks.
So it treats the isospin and the strangeness in a different
manner, i.e., rotational and vibrational, respectively. In this model,
hyperons are described by bound states of kaon(s) in an SU(2) soliton
background field. The resulting mass spectrum \cite{CK,BRS,SMNR},
electromagnetic and axial properties of hyperons \cite{EM,weak} are
consistent with the existing experimental data, so this model may be a
reasonable starting point for understanding the hyperon structure. It is
also shown \cite{OMRS,OP} that the magnetic moments of baryons have the
same structure of the quark model predictions. So, it will be
interesting to investigate other electromagnetic properties of hyperons
in this model. In addition, this model predicts small strangeness
content of the proton, $\langle s\bar s \rangle_p \approx 3$-6\%
\cite{BRS,WVC}. Therefore, it will also be interesting to test the argument
of Ref. \cite{KS} on the relation between the electric quadrupole moments and
the SU(3) symmetry breaking.

We start with the effective action \cite{OMRS,RS} for the Skyrme model
with proper symmetry breaking terms, which reads
\begin{eqnarray}
\Gamma &=& \int d^4 x \Bigl\{ \frac{1}{16} F_\pi^2 \mbox{Tr}\,
         ( \partial_\mu U \partial^\mu U^\dagger )
\nonumber \\ && \mbox{}
       + \frac{1}{32e^2} \mbox{Tr}\,
         [ U^\dagger \partial_\mu U, U^\dagger \partial_\nu U ]^2
         \Bigr\}
       + \Gamma_{\text{WZ}}
       + \Gamma_{\text{SB}},
\end{eqnarray}
where $\Gamma_{\text{WZ}}$ is the Wess-Zumino action and
$\Gamma_{\text{SB}}$ includes the fact that the pion decay
constant $F_\pi$ ($\approx 186$ MeV empirically) is not equal to the
kaon decay constant $F_K$ ($\approx 1.22 F_\pi$ experimentally) as
well as the meson mass terms:
\begin{eqnarray}
\Gamma_{\text{SB}} &=& \int d^4 x \Bigl\{
       \frac{F_\pi^2}{48} ( m_\pi^2 + 2 m_K^2 ) \, \mbox{Tr}\,
           ( U+U^\dagger-2 )
\nonumber \\ && \mbox{}
     + \frac{\sqrt3}{24} F_\pi^2 ( m_\pi^2 - m_K^2 ) \mbox{Tr}\,
           [ \lambda_8 (U + U^\dagger) ]
\nonumber \\ && \mbox{}
+ \frac{F_K^2 - F_\pi^2}{48}\, \mbox{Tr}\,
        \{ ( 1 - \sqrt3 \lambda_8 )
               [ 2 m_K^2 ( U + U^\dagger - 2 )
\nonumber \\ && \qquad \qquad \mbox{}
               + ( U \partial_\mu U^\dagger \partial^\mu U
               + U^\dagger \partial_\mu U \partial^\mu U^\dagger )
  ] \} \Bigr\},
\end{eqnarray}
with the eighth Gell-Mann matrix $\lambda_8$, where $m_\pi$ and $m_K$
are the pion and kaon mass, respectively. Taking the Callan-Klebanov
ansatz \cite{CK}, the SU(3)-valued chiral field $U$ is written as
\begin{eqnarray}
U =
  \left( \begin{array}{cc}
         \exp( i \bbox{\tau} \cdot \bbox{\pi} / F_\pi ) & 0 \\
         0 & 1 \end{array} \right)
  \exp \left[ \frac{i 2 \sqrt2}{F_\pi}
       \left( \begin{array}{cc} 0 & K \\
              K^\dagger & 0 \end{array} \right) \right]
 \left( \begin{array}{cc}
        \exp( i \bbox{\tau} \cdot \bbox{\pi} / F_\pi ) & 0 \\
        0 & 1 \end{array} \right).
\end{eqnarray}
By expanding the action up to the second order of $K$, one can obtain
the kaon-soliton effective Lagrangian. It is well-known that this
effective Lagrangian with the hedgehog configuration $\exp(2i
\bbox{\tau} \cdot \bbox{\pi} /F_\pi) = \exp[iF(r) \bbox{\tau} \cdot
\hat{\bf r}]$ supports a bound state solution of the lowest
kaon state in the form of $K = e^{i\omega t}k(r) \bbox{\tau} \cdot
\hat{\bf r} \chi$, where $\chi$ is a two-component isospinor.
(See Refs. \cite{CK,SMNR} for details.)

With the given Lagrangian, we can construct the electromagnetic charge
density, which reads

\begin{equation}
\rho_e ({\bf r}) = {\textstyle \frac12} Y ({\bf r}) + I^3 ({\bf r}),
\label{charge}
\end{equation}
where
\begin{eqnarray}
Y({\bf r}) &=& - \frac{1}{2\pi^2} \frac{\sin^2 F}{r^2} F' |k|^2
               - 2 [ f(r) \omega + \lambda(r) ] |k|^2, \nonumber \\
I^3 ({\bf r}) &=& V(r) ( \delta^{bc}
                - \hat{\bf r}^b \hat{\bf r}^c) D^{3c} \Omega^b
\nonumber \\
&& \mbox{} + [ A(r) \tau^b - B(r) \hat{\bf r}^b
            \bbox{\tau} \cdot \hat{\bf r} ] D^{3b}.
\label{YI}
\end{eqnarray}
(We keep the conventions of Refs. \cite{OMRS,RS} throughout this work.)
The functionals $f(r)$, $\lambda(r)$, $V(r)$, $A(r)$, and $B(r)$ are
explicitly
\begin{eqnarray}
f(r) &=& 1 + \frac{1}{e^2 F_\pi^2 \chi^2}
         \left( F'^2 + 2 \frac{\sin^2 F}{r^2} \right), \nonumber \\
\lambda (r) &=& - \frac{N_c}{2\pi^2 F_\pi^2 \chi^2}
         \frac{\sin^2 F}{r^2} F', \nonumber \\
V(r) &=& -2 \sin^2 F \left\{ \frac{F_\pi^2}{4}
           + \frac{1}{e^2} \left( F'^2 + \frac{\sin^2 F}{r^2} \right)
           \right\}, \nonumber \\
A(r) &=& \omega |k|^2 f(r) \cos F - |k|^2 \lambda(r) \nonumber \\
     && \mbox{} + \frac{4\omega}{e^2F_\pi^2\chi^2}
          \{ |k|^2 \frac{\sin^2 F}{r^2}
          \cos^2 \frac{F}{2} + \frac32 kk' \sin F F'
          \}, \nonumber\\
B(r) &=& \omega |k|^2 f(r) (1 + \cos F ) \nonumber \\
     && \mbox{} + \frac{4\omega}{e^2F_\pi^2\chi^2}
          \{ |k|^2 \frac{\sin^2 F}{r^2}
          \cos^2 \frac{F}{2} + \frac32 kk' \sin F F' \},
\nonumber \\
\end{eqnarray}
where $N_c$ is the number of color and $\chi = F_K / F_\pi$. In terms of
the collective rotation variable ${\cal A}(t)$, $D^{ab} = \frac12 \mbox{Tr}
( \tau^a {\cal A} \tau^b {\cal A}^\dagger )$ and $\Omega^a = - \frac{i}{2}
\mbox{Tr} ( \tau^a {\cal A}^\dagger \dot {\cal A} )$. We take the physical
pion and kaon mass, $m_\pi = 138$ MeV and $m_K = 495$ MeV, while the pion
decay constant $F_\pi$ and the Skyrme parameter $e$ are chosen to be $F_\pi =
108$ MeV and $e=4.84$ \cite{AN} to reproduce the nucleon and $\Delta$ masses.
Then the energy of the bound kaon $\omega$ is calculated as 146 MeV and 209 MeV
for $\chi=1.0$ and 1.22, respectively \cite{OMRS}.

The electric quadrupole moment operator is defined by
\begin{equation}
\hat Q_{ij} = \int {\rm d}^3 r
              \left( r_i r_j - {\textstyle \frac13}
              r^2 \delta_{ij} \right) \rho_e ({\bf r}),
\end{equation}
and the quadrupole moment of a baryon $Q_{33} (B) = \langle B | \hat
Q_{33} | B \rangle $ can be read from the baryon wave function
\cite{OMRS}. From Eq. (\ref{charge}), it is straightforward to obtain
the form of $Q_{33}$ as
\begin{eqnarray}
\hat Q_{33} &=& - a_1 ( D^{3b} R^b - 3 D^{33} R^3 )
\nonumber \\ && \mbox{}
           - ( ca_1 + a_2 ) (D^{3b} J_k^b - 3 D^{33} J_k^3 ),
\label{QE2}
\end{eqnarray}
where
\begin{eqnarray}
a_1 &=& - \frac{4\pi}{45{\cal I}} \int {\rm d}r \, r^4 V(r) =
{\textstyle\frac{1}{15}} \langle r^2 \rangle^N_{I=1}, \nonumber\\
a_2 &=& \frac{4\pi}{45} \int {\rm d} r \, r^4 B(r).
\label{a12}
\end{eqnarray}
For the diagonal matrix elements one has $D^{3i} = - \frac{I^3
R^i}{I(I+1)}$, and the formula (\ref{QE2}) can be more simplified as
\begin{equation}
\hat Q_{33} (B) = \sum_{k=1}^2 a_k \hat {\cal O}_k \frac{I_3}{I(I+1)},
\label{QE2d}
\end{equation}
where
\begin{eqnarray}
\hat {\cal O}_1 &=& I(I+1) - 3 R_3^2 + c \, \hat {\cal O}_2, \nonumber
\\
\hat {\cal O}_2 &=& {\textstyle\frac12} \left\{ J(J+1) - I(I+1) - J_k
(J_k+1) \right\} - 3 J_{k,3} R_3.
\nonumber \\
\label{O12}
\end{eqnarray}
Here, ${\cal I}$ is the moment of inertia of the SU(2) soliton and
$\langle r^2 \rangle^N_{I=1}$ is the isovector mean square radius of
the nucleon. Therefore, for nonstrange baryons the formula (\ref{QE2})
reproduces the SU(2) model predictions \cite{WW,Adkins}. In Eq. (\ref{O12}),
$I$ ($J$) is the isospin (spin) of the baryon, and $J_k$ and $R$ are
the kaon grand spin and the rotor spin, respectively \cite{CK,SMNR}.
The hyperfine constant $c$ is obtained as $c=0.51$ and 0.39 for
$\chi=1.0$ and 1.22, respectively.

{}From Eq. (\ref{YI}), one can see that only the isovector current
contributes to the quadrupole moments since the hypercharge density is
spherically symmetric. It also should be noted that the $Q_{33}$
depends on the isospin of the baryon. As noted in Refs. \cite{LDW,KS},
this dependence is not consistent with the quark model predictions
where the quadrupole moments depend on the baryon charge. This dependence
supports the argument of Ref. \cite{KS} when we consider the small
strangeness content of the proton of this model.
Furthermore, the $Q_{33}$ contains a kaonic contribution, i.e.,
the $a_2$ term, of which effect is comparable to the pionic one, whereas in
the slow rotator approach of Ref. \cite{KS}, it depends only on the
pionic current because its moment of inertia is spherical
outside the SU(2) subspace. Numerically, we have $a_1 = 0.0735$ fm$^2$ and
$a_2 = 0.0528$ (0.0397) fm$^2$ for $\chi=1.0$ (1.22).

The transitional electric quadrupole moments of the decuplet baryons
can be obtained from Eq. (\ref{QE2}). The expectation values
of the operator $D^{ab}$ can be evaluated by making use of the formula
\cite{Adkins}
\begin{eqnarray}
&& \langle I, I_3, R_3 | D^{ab} | I', I_3', R_3' \rangle = \left[
\frac{2I'+1}{2I+1} \right]^{1/2} (-1)^{I-I_3 - I'+I_3'}
\nonumber \\
&& \qquad \times
\langle I' \, -I_3' \, ; 1 \, a \, | \, I \, -I_3 \rangle
\langle I' \, R_3' \, ; 1 \, b \, | \, I \, R_3 \rangle.
\end{eqnarray}
The off-diagonal matrix elements of the operators $D^{33} R^3$,
$D^{33} R^b$, $D^{33} J_k^3$, and $D^{3b} J_k^b$ are given in Table I.

Our results are summarized in Tables II and III. Given in Table II are
the predictions for the electric quadrupole moments of the decuplet
baryons. For a comparison, the predictions of other models are also
presented. The transitional electric quadrupole moments are listed
in Table III. In quark models, the electric quadrupole moments are
proportional to the baryon charge and the nonvanishing quadrupole
moments of neutral hyperons come from the quark mass difference
$m_{u,d} \neq m_s$. But this dependence should be corrected if one
includes meson cloud effects as discussed in Ref. \cite{LG}. Then
the quadrupole moments are nearly isospin-dependent rather than
charge-dependent, which is consistent with our results. Also, our
results are closer to the predictions of the heavy baryon chiral
perturbation theory \cite{BSS} than to those of the modified SU(3)
collective model \cite{KS}.

As a test of tensor force in the quark model, it has been suggested
to measure the electric quadrupole moment of the $\Omega^-$ because of
its long life-time. In Ref. \cite{SG}, it was suggested to
measure the quadrupole moment of $\Omega^-$ by studying the hyperfine
splitting of the exotic $\Omega^-$-nucleus atoms if $Q_{33}(\Omega^-)$
is larger than 0.01 e$\cdot$fm$^2$ \cite{Ric,KG}. However, if the quadrupole
moments have isospin dependence it may be difficult to measure the
$Q_{33} (\Omega^-)$ experimentally.

As a summary, we have calculated the static and transitional electric
quadrupole moments of the decuplet baryons in the bound state approach
of the Skyrme model. The results show the isospin dependence of the
quadrupole moments. It would hence be of great interest if such quantities
could be measured experimentally, which will give very useful
informations on the baryon structure.

\vskip 0.5cm
We are grateful to Dong-Pil Min and Gwang-Ho Kim for useful
discussions at the early stage of this work. Encouragements from Shin Nan
Yang are also gratefully acknowledged. This research was
supported in part by the National Science Council of ROC under Grant
No. NSC84-2811-M002-036.

\newpage
\widetext

\begin{table}
\begin{center}
\begin{tabular}{l|l}
$\langle p, J_z = {\textstyle\frac12} | D^{33} R^3 | \Delta^+, J_z =
{\textstyle\frac12} \rangle = {\textstyle - \frac{\sqrt2}{6}}$ &
$\langle p, J_z = {\textstyle\frac12} | D^{3b} R^b | \Delta^+, J_z =
{\textstyle\frac12} \rangle = {\textstyle 0}$  \\
$\langle \Lambda, J_z = {\textstyle\frac12} | D^{33} R^3 | \Sigma^0, J_z =
{\textstyle\frac12} \rangle = {\textstyle 0}$ &
$\langle \Lambda, J_z = {\textstyle\frac12} | D^{3b} R^b | \Sigma^0, J_z =
{\textstyle\frac12} \rangle = {\textstyle 0}$  \\
$\langle \Lambda, J_z = {\textstyle\frac12} | D^{33} R^3 | \Sigma^{*,0}, J_z =
{\textstyle\frac12} \rangle = {\textstyle 0}$ &
$\langle \Lambda, J_z = {\textstyle\frac12} | D^{3b} R^b | \Sigma^{*,0}, J_z =
{\textstyle\frac12} \rangle = {\textstyle 0}$  \\
$\langle \Sigma^a, J_z = {\textstyle\frac12} | D^{33} R^3 | \Sigma^{*,a}, J_z =
{\textstyle\frac12} \rangle = {\textstyle - \frac{\sqrt2}{6} a}$ &
$\langle \Sigma^a, J_z = {\textstyle\frac12} | D^{3b} R^b | \Sigma^{*,a}, J_z =
{\textstyle\frac12} \rangle = {\textstyle - \frac{\sqrt2}{3} a}$  \\
$\langle \Xi^a, J_z = {\textstyle\frac12} | D^{33} J_k^3 | \Xi^{*,a}, J_z =
{\textstyle\frac12} \rangle = {\textstyle 0}$ &
$\langle \Xi^a, J_z = {\textstyle\frac12} | D^{3b} J_k^b | \Xi^{*,a}, J_z =
{\textstyle\frac12} \rangle = {\textstyle 0}$  \\
\hline
$\langle p, J_z = {\textstyle\frac12} | D^{33} J_k^3 | \Delta^+, J_z =
{\textstyle\frac12} \rangle = {\textstyle 0}$ &
$\langle p, J_z = {\textstyle\frac12} | D^{3b} J_k^b | \Delta^+, J_z =
{\textstyle\frac12} \rangle = {\textstyle 0}$  \\
$\langle \Lambda, J_z = {\textstyle\frac12} | D^{33} J_k^3 | \Sigma^0, J_z =
{\textstyle\frac12} \rangle = {\textstyle \frac16}$ &
$\langle \Lambda, J_z = {\textstyle\frac12} | D^{3b} J_k^b | \Sigma^0, J_z =
{\textstyle\frac12} \rangle = {\textstyle \frac12}$  \\
$\langle \Lambda, J_z = {\textstyle\frac12} | D^{33} J_k^3 | \Sigma^{*,0}, J_z
=
{\textstyle\frac12} \rangle = {\textstyle -\frac{\sqrt2}{6}}$ &
$\langle \Lambda, J_z = {\textstyle\frac12} | D^{3b} J_k^b | \Sigma^{*,0}, J_z
=
{\textstyle\frac12} \rangle = {\textstyle 0}$ \\
$\langle \Sigma^a, J_z = {\textstyle\frac12} | D^{33} J_k^3 | \Sigma^{*,a},
J_z = {\textstyle\frac12} \rangle = {\textstyle  \frac{\sqrt2}{12} a}$ &
$\langle \Sigma^a, J_z = {\textstyle\frac12} | D^{3b} J_k^b | \Sigma^{*,a},
J_z = {\textstyle\frac12} \rangle = {\textstyle \frac{\sqrt2}{6} a}$  \\
$\langle \Xi^a, J_z = {\textstyle\frac12} | D^{33} J_k^3 | \Xi^{*,a}, J_z =
{\textstyle\frac12} \rangle = {\textstyle  \frac{2\sqrt2}{9} (a+\frac12)}$ &
$\langle \Xi^a, J_z = {\textstyle\frac12} | D^{3b} J_k^b | \Xi^{*,a}, J_z =
{\textstyle\frac12} \rangle = {\textstyle \frac{2(\sqrt2-1)}{9} (a+\frac12)}$\\
\end{tabular}
\end{center}
\caption{Off-diagonal matrix elements of the operators in Eq.
(\protect\ref{QE2}).}
\end{table}

\begin{table}
\begin{center}
\begin{tabular}{c|r|r|r|r|r|r|r|r|r|r}
Particle & $\Delta^{++}$ & $\Delta^+$ & $\Delta^0$ & $\Delta^-$ &
       $\Sigma^{*+}$ & $\Sigma^{*0}$ & $\Sigma^{*-}$ &
       $\Xi^{*0}$ & $\Xi^{*-}$ & $\Omega^-$ \\ \hline
Ref. \cite{GZ} & & & & & & & & & & 1.8 \\
Ref. \cite{GD} & $-6.6$ & $-3.3$ & $0.0$ & $3.3$ & & & & & &  \\
Ref. \cite{IKK} & $-9.8$ & $-4.9$ & $0.0$ & $4.9$ & & & & & & 3.1 \\
Ref. \cite{Ric} & & & & & & & & & & 0.4 \\
Ref. \cite{DG} & $-17.8$ & $-8.9$ & $0.0$ & $8.9$ & & & & & &  \\
Ref. \cite{Kri} & $-12.6$ & $-6.3$ & $0.0$ & $6.3$ & & & & & &  \\
Ref. \cite{LG} & $-6.0$ & $-2.1$ & $1.8$ & $5.7$ & $-2.2$ & $-0.01$ &
2.0 & $-0.6$ & 1.0 & 0.6 \\
Ref. \cite{KG} & $-9.3$ & $-4.6$ & $0.0$ & 4.6 & $-5.4$ & $-0.7$ & 4.0
& $-1.3$ &3.4 & 2.8 \\
Ref. \cite{LDW} & $-2.7$ & $-1.3$ & 0.0 & 1.3 & 0.2 & 0.5 & 1.0 & 0.5 &
0.8 & 0.5 \\
Ref. \cite{BSS} & $-8.0$ & $-3.0$ & 1.2 & 6.0 & $-7.0$ & $-1.3$ & 4.0 &
$-3.5$ & 2.0 & 0.9 \\
Ref. \cite{KS} & $-8.7$ & $-3.1$ & 2.4 & 8.0 & $-4.2$ & 0.5 & 5.2 &
$-0.7$ & 3.5 & 2.4 \\
This Work (I) & $-8.8$ & $-2.9$ & 2.9 & 8.8 & $-8.2$ & 0.0 & 8.2 &
$-6.0$ & 6.0 & 0.0 \\
This Work (II) & $-8.8$ & $-2.9$ & 2.9 & 8.8 & $-7.1$ & 0.0 & 7.1 &
$-4.6$ & 4.6 & 0.0 \\
\end{tabular}
\end{center}
\caption{Electric quadrupole moments of the decuplet baryons in the
unit of $10^{-2}$ e$\cdot$fm$^2$. The work (I) and (II) correspond to
$\chi=1.0$ and 1.22, respectively.}
\end{table}

\begin{table}
\begin{center}
\begin{tabular}{c|c|c|c|c|c|c|c}
$\Delta^+ \to p$ & $\Delta^0 \to n$ & $\Sigma^{*,0} \to \Lambda$
& $\Sigma^{*,+} \to \Sigma^+$ & $\Sigma^{*,0} \to \Sigma^0$
& $\Sigma^{*,-} \to \Sigma^-$ & $\Xi^{*,0} \to \Xi^0$ &
$\Xi^{*,-} \to \Xi^-$ \\ \hline
$-5.20$ & 5.20 & $-4.83$ & $-0.93$ & 0 & 0.93 & 2.91 & $-2.91$ \\
\end{tabular}
\end{center}
\caption{Transitional electric quadrupole moments of the decuplet
baryons with $\chi=1.22$ in the unit of $10^{-2}$ e$\cdot$fm$^2$.}
\end{table}

\begin{references}

\bibitem{DGG}
   A. De R\'ujula, H. Georgi, and S. L. Glashow,
        Phys. Rev. D {\bf 12}, 147 (1975).

\bibitem{BM}
   C. Becchi and G. Morpurgo,
        Phys. Lett. {\bf 17}, 352 (1965).

\bibitem{GZ}
   S. S. Gershte\u{\i}n and Yu. M. Zinov'ev,
        Yad. Fiz. {\bf 33}, 1442 (1981)
        [Sov. J. Nucl. Phys. {\bf 33}, 772 (1981)].

\bibitem{GD}
   S. S. Gershte\u{\i}n and G. V. Dzhikiya,
        Yad. Fiz. {\bf 34}, 1566 (1981)
        [Sov. J. Nucl. Phys. {\bf 34}, 870 (1981)].

\bibitem{IKK}
   N. Isgur, G. Karl, and R. Koniuk,
        Phys. Rev. D {\bf 25}, 2394 (1982).

\bibitem{Ric}
   J.-M. Richard,
        Z. Phys. C {\bf 12}, 369 (1982).

\bibitem{DG}
   D. Drechsel and M. M. Giannini,
        Phys. Lett. B {\bf 143}, 329 (1984).

\bibitem{Kri}
   M. I. Krivoruchenko,
        Yad. Fiz. {\bf 45}, 809 (1987)
        [Sov. J. Nucl. Phys. {\bf 45}, 503 (1987)];
        Z. Phys. C {\bf 36}, 243 (1987).

\bibitem{LG}
   W. J. Leonard and W. J. Gerace,
        Phys. Rev. D {\bf 41}, 924 (1990).

\bibitem{KG}
   M. I. Krivoruchenko and M. M. Giannini,
        Phys. Rev. D {\bf 43}, 3763 (1991).

\bibitem{LDW}
   D. B. Leinweber, T. Draper, and R. M. Woloshyn,
        Phys. Rev. D {\bf 46}, 3067 (1992).

\bibitem{BSS}
   M. N. Butler, M. J. Savage, and R. P. Springer,
        Phys. Rev. D {\bf 49}, 3459 (1994).

\bibitem{KS}
   J. Kroll and B. Schwesinger,
        Phys. Lett. B {\bf 334}, 287 (1994).

\bibitem{JM}
   E. Jenkins and A. V. Manohar,
        Phys. Lett. B {\bf 255}, 558 (1991).

\bibitem{SW}
   B. Schwesinger and H. Weigel,
       Nucl. Phys. {\bf A450}, 461 (1992).

\bibitem{CK}
   C. G. Callan and I. Klebanov,
        Nucl. Phys. {\bf B262}, 365 (1985);
   C. G. Callan, K. Hornbostel, and I. Klebanov,
        Phys. Lett. B {\bf 202}, 269 (1988).

\bibitem{BRS}
   J.-P. Blaizot, M. Rho, and N. N. Scoccola,
        Phys. Lett. B {\bf 209}, 27 (1988).

\bibitem{SMNR}
   N. N. Scoccola, D.-P. Min, H. Nadeau, and M. Rho,
        Nucl. Phys. {\bf A505}, 497 (1989).

\bibitem{EM}
   J. Kunz and P. J. Mulders,
        Phys. Lett. B {\bf 231}, 335 (1989);
        Phys. Rev. D {\bf 41}, 1578 (1990);
   E. M. Nyman and D. O. Riska,
        Nucl. Phys. {\bf B325}, 593 (1989);
   D.-P. Min, Y. S. Koh, Y. Oh, and H. K. Lee,
        Nucl. Phys. {\bf A530}, 698 (1991).

\bibitem{weak}
   Y. Kondo, S. Saito, and T. Otofuji,
        Phys. Lett. B {\bf 236}, 1 (1990);
        {\bf 256}, 316 (1991);
   Y. Kondo and S. Saito,
        Few-Body Syst. {\bf 12}, 113 (1992);
   K. Dannbom and D. O. Riska,
        Nucl. Phys. {\bf A548}, 669 (1992).

\bibitem{OMRS}
   Y. Oh, D.-P. Min, M. Rho, and N. N. Scoccola,
        Nucl. Phys. {\bf A534}, 493 (1991).

\bibitem{OP}
   Y. Oh and B.-Y. Park
        (in preparation).

\bibitem{WVC}
   C. W. Wong, D. Vuong, and K.-C. Chu,
        Nucl. Phys. {\bf A515}, 686 (1990).

\bibitem{RS}
   D. O. Riska and N. N. Scoccola,
        Phys. Lett. B {\bf 265}, 188 (1991).

\bibitem{AN}
   G. S. Adkins and C. R. Nappi,
        Nucl. Phys. {\bf B233}, 109 (1984).

\bibitem{WW}
   A. Wirzba and W. Weise,
        Phys. Lett. B {\bf 188}, 6 (1987).

\bibitem{Adkins}
   G. S. Adkins,
        in {\it Chiral Solitons\/}, edited by K.-F. Liu
        (World Scientific, Singapore, 1987).

\bibitem{SG}
   R. M. Sternheimer and M. Goldhaber,
        Phys. Rev. A {\bf 8}, 2207 (1973).

\end{references}
\end{document}